\documentclass[pdflatex,aps,prl,nofootinbib,twocolumn,showpacs,superscriptaddress]{revtex4-2}
\usepackage{amssymb,amsmath,color,hyperref,graphicx,multirow,tabularx,physics}

\begin{document}


\title{Chiral Bosonic Topological Insulator on the Honeycomb Lattice with Anisotropic Interactions} 


\author{Amrita Ghosh}
\author{Eytan Grosfeld}
\affiliation{Department of Physics, Ben-Gurion University of the Negev, Beer-Sheva 8410501, Israel}

\date{\today}


\begin{abstract}

We study hard-core bosons on the honeycomb lattice in the presence of anisotropic nearest-neighbor repulsive interactions. Using a quantum Monte Carlo (QMC) technique, we extract the phase diagram of the model 
in terms of the filling and the anisotropy. At half-filling we find a dimer insulator phase near maximum anisotropy that is characterized by a finite topological entanglement entropy $\ln(2)/2$, indicative of a fractional quantum Hall state for bosons.
We identify the presence of edge states and derive a QMC-based method to extract and verify their chirality. Remarkably, this phase arises in the absence of magnetic flux and without explicit lattice frustration.

\end{abstract}

\maketitle


\emph{Introduction.---} Topological phases of electrons are ubiquitous in condensed matter physics. Interest in these states started following the discovery of the quantum Hall effect (QHE)~\cite{klitzing1980new,tsui1982two}, which can form in extremely low temperatures and high magnetic fields. Later, Haldane proposed a model on the honeycomb lattice demonstrating the possibility for the formation of a QHE even when the total magnetic field is zero~\cite{haldane1988model}. This opened a pathway towards the prediction and consequent experimental realization of modern topological insulators, including the two-dimensional (2D) spin-Hall effect~\cite{bernevig2006quantum} and the three-dimensional topological insulators~\cite{fu2007topological,moore2007topological,roy2009topological}. Unlike their forefather, these states can form even in time-reversal symmetric situations and in relatively high temperatures.

When the underlying particles are bosons, the realization of topological phases requires the presence of interactions that act to stabilize these phases. The presence of interactions can further lead to richer, interacting topological phases, among which the realization of bosonic quantum Hall states was suggested in either large magnetic fields (or rotation) \cite{cooper2008rapidly,gerster2017fractional} or in frustrated lattices \cite{kalmeyer1987equivalence,gong2014emergent} or by explicitly breaking time reversal symmetry using a local flux \cite{wang2011fractional,nielsen2013local}. Simpler models realizing the QHE for bosons could constitute crucial building blocks towards the modeling and realization of additional bosonic topological phases. 

In this paper, we consider hard-core bosons (HCBs) on a 2D honeycomb lattice subjected to anisotropic bond interactions. Using Stochastic Series Expansion (SSE) technique \cite{sandvik1997finite,sandvik2010computational}, a quantum Monte Carlo (QMC) method, we extract the full phase diagram of the model. In the isotropic limit our model coincides with the popular $t-V$ model on the 2D honeycomb lattice, whose phase diagram has been extensively studied long ago \cite{stefan2007phase}. It consists of a charge-density-wave (CDW) phase at half-filling which phase separates into a superfluid phase as the chemical potential of the system is tuned. We demonstrate that as we increase the anisotropy, the CDW insulator at $\rho=1/2$ transforms into a time reversal breaking topological insulator, which admits a finite topological entanglement entropy (TEE) of value $\ln(2)/2$ and chiral edge states. We therefore argue that this phase is consistent with the fractional quantum Hall liquid for bosons at filling factor $1/2$ \cite{haque2007entanglement}. 
Exceptionally, the phase is realized on the non-frustrated honeycomb lattice in the absence of any type of flux. 

\begin{figure}[b]
	\centering
	\includegraphics[width=0.45\textwidth]{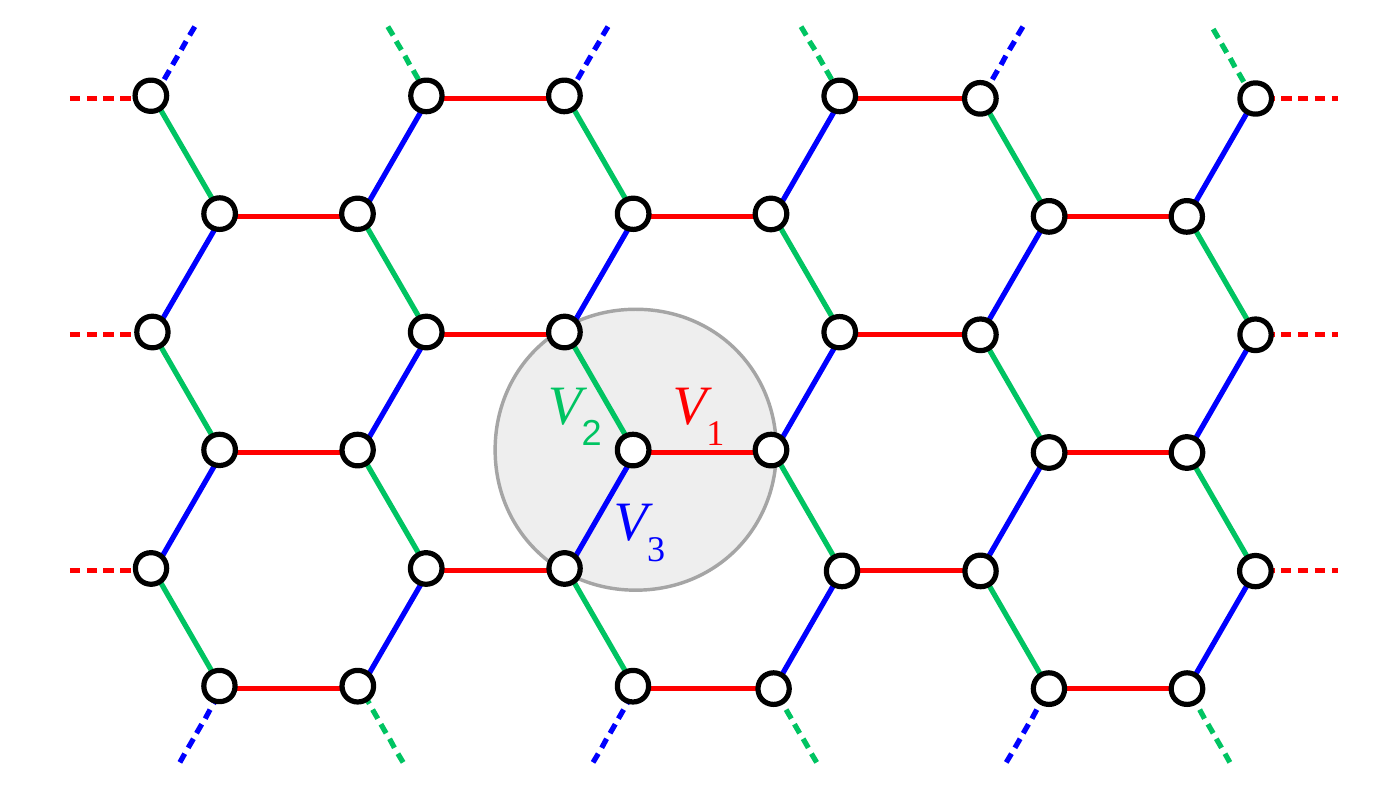}
	\caption{Schematic of the model described in the main text for hard-core bosons on the honeycomb lattice. The bonds represent hopping of strength $t$ and nearest-neighbor repulsive interactions of magnitude $V_1$ (red), $V_2$ (green) and $V_3$ (blue).}
	\label{fig:model}
\end{figure}


\begin{figure*}[t]
	\centering
	\includegraphics[width=0.343\textwidth]{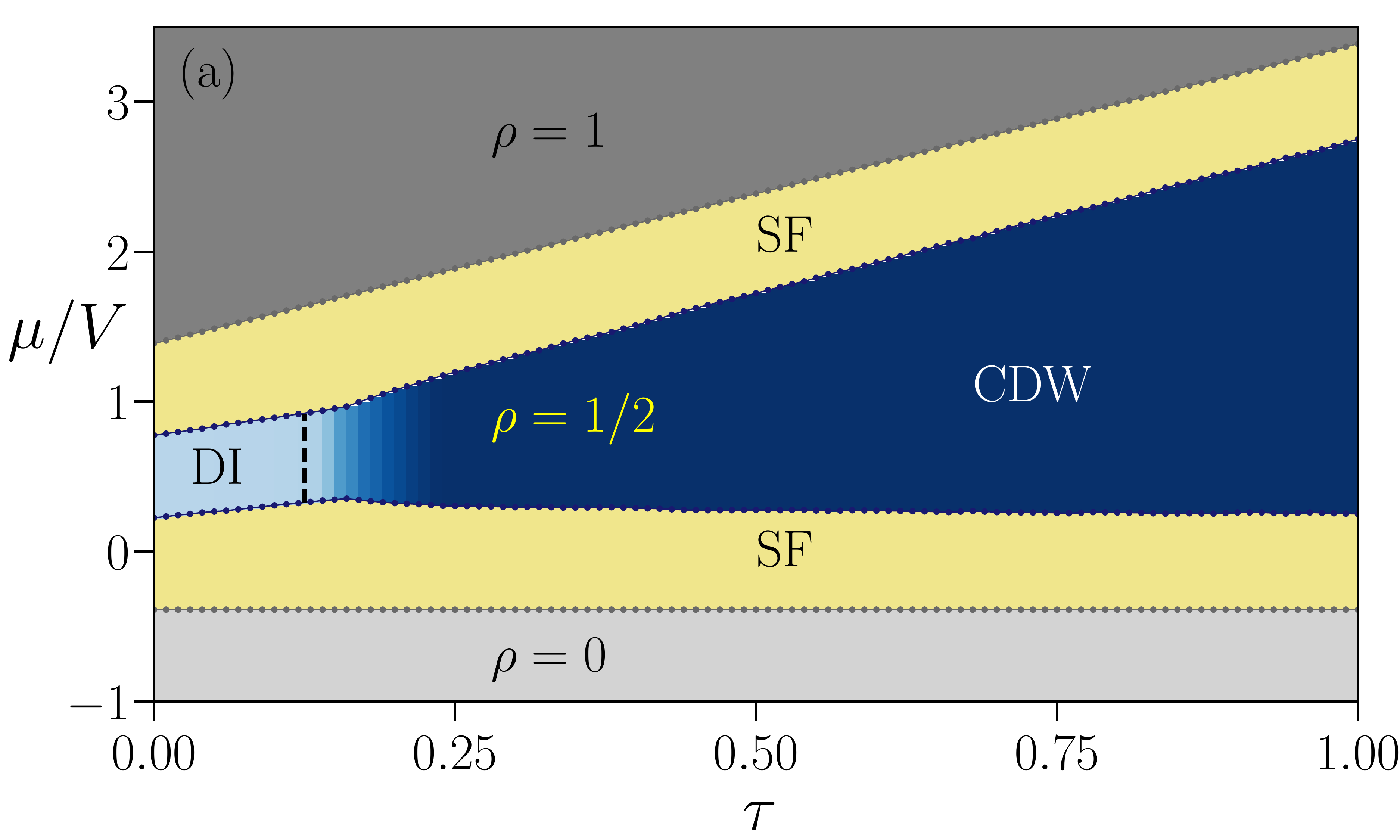}
	\includegraphics[width=0.30\textwidth]{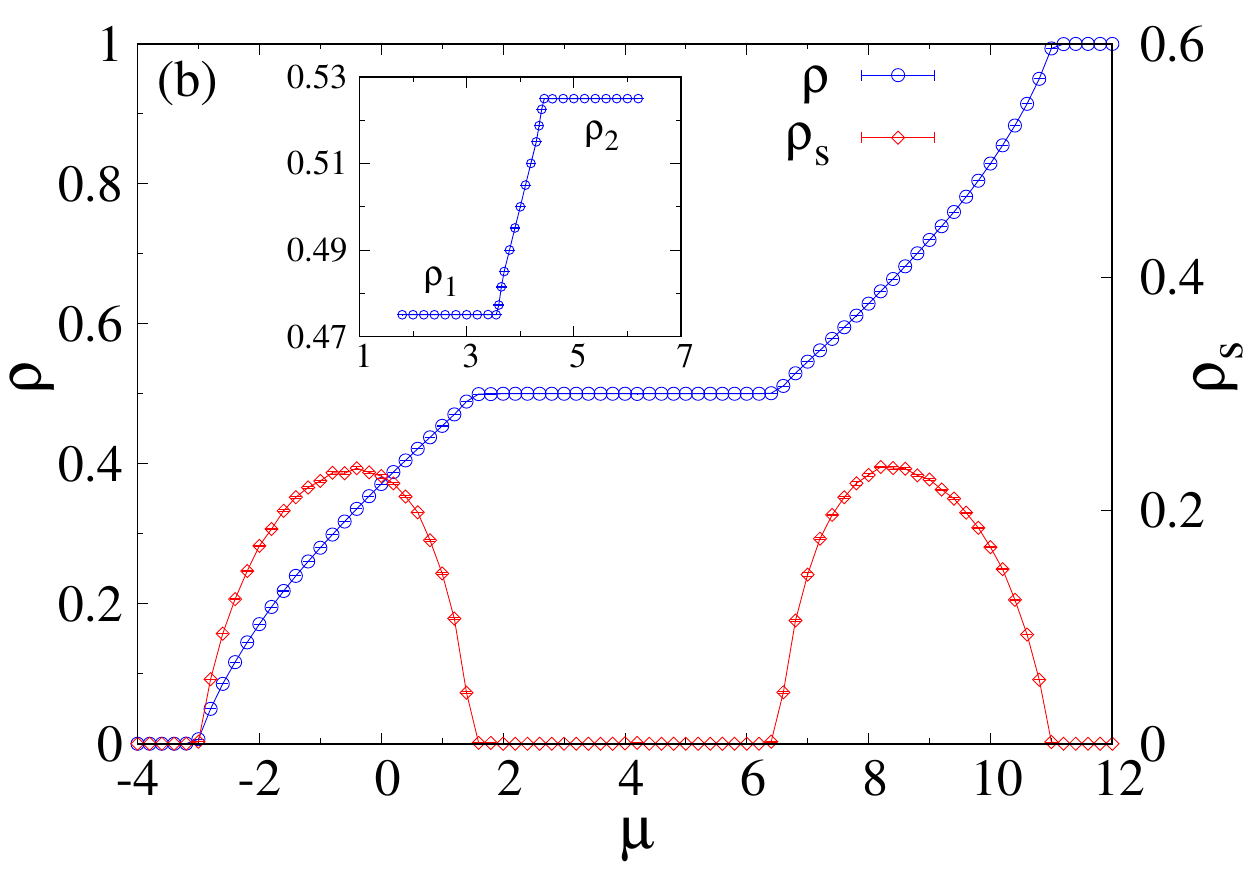}
	\includegraphics[width=0.30\textwidth]{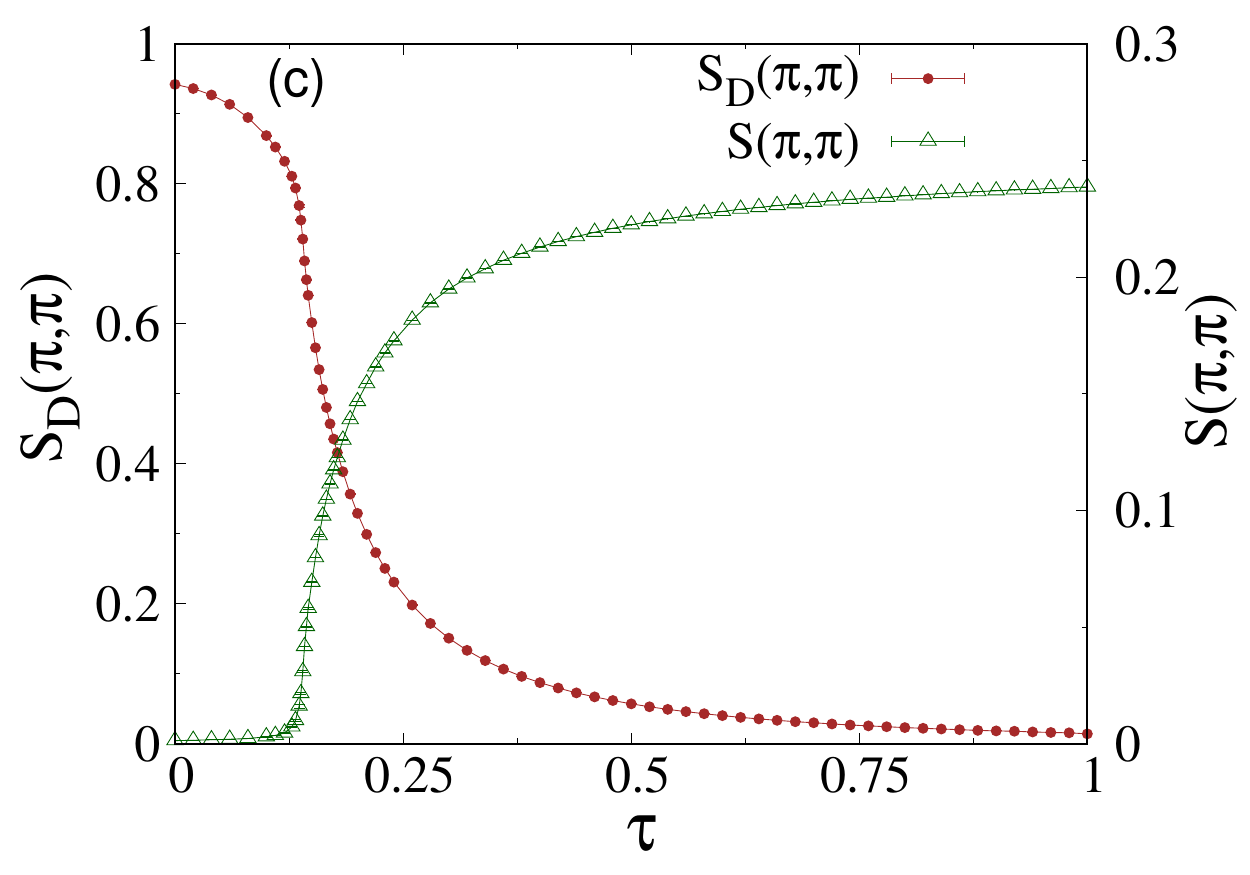}
	\caption{Phase diagram and order parameters. (a) Complete phase diagram corresponding to the Hamiltonian in Eq.~\eqref{eq:hamiltonian} in terms of $\mu/V$ and $\tau$ for a $20\times20$ periodic honeycomb lattice with $V=8$. Variation of (b) average density $\rho$ $\&$ superfluid density $\rho_s$ versus $\mu$, and (c) structure factor $S(\pi,\pi)$ $\&$ dimer structure factor $S_D(\pi,\pi)$ versus $\tau$, on a $20\times 20$ periodic honeycomb lattice with $V=8$, $V^\prime=0$ and $t=1$. Inset of (b): Splitting of $\rho = 1/2$ plateau under a straight cut parallel to the $y$-axis.}
	\label{fig:phase_diagram}
\end{figure*}

\emph{Model and phase diagram.---}
In our model the HCBs are subjected to anisotropic repulsive interactions on a 2D Honeycomb lattice according to the Hamiltonian,
\begin{align}
 \hat{H}=-t&\sum\limits_{\langle i,j\rangle} \left(\hat{d}_i^\dagger \hat{d}_j+\rm{h.c.}\right)+\sum_{\alpha=1}^3 V_\alpha\sum\limits_{\langle i,j\rangle_\alpha}\hat{n}_i \hat{n}_j -\mu\sum\limits_i \hat{n}_i \label{eq:hamiltonian}
\end{align}
where $\hat{d}_i^\dagger$ ($\hat{d}_i$) is the creation (annihilation) operator of a HCB at site $i$, $\hat{n}_i=\hat{d}_i^\dagger \hat{d}_i$ is the density at the same site, $t$ is the hopping amplitude along nearest-neighbor (NN) bonds $\langle i,j \rangle$ and $\mu$ is the chemical potential. 
The HCBs experience NN repulsion $V_\alpha$ on bonds $\langle i,j\rangle_\alpha$ which belong to one of the three families $\alpha$ of parallel bonds highlighted in Fig.~\ref{fig:model}\,. For convenience, in the following we take  $\boldsymbol{V}=(V,V',V')$; the parameter $\tau=V'/V$ is therefore a measure of the isotropy, with $\tau=1$ indicating full isotropy and $\tau=0$ maximal anisotropy. We perform the simulations using $N_x\times N_y=N_s$ sites. For the definitions of the various order parameters that we use to characterize the Hamiltonian we refer the reader to \cite{ghosh2020weak}.

The complete phase diagram corresponding to Eq.~\eqref{eq:hamiltonian} is shown in Fig.~\ref{fig:phase_diagram}\,a in the $(\mu/V,\tau)$ plane. Apart from the usual empty phase at density $\rho=0$ and Mott insulator at $\rho=1$, the phase diagram basically consists of an insulating phase at density $1/2$ surrounded by a superfluid phase. Fig.~\ref{fig:phase_diagram}\,b displays the dependence of the average HCB density $\rho$ and the superfluid density $\rho_s$ on the chemical potential of the system $\mu$ at $\tau=0$. The plateau in the $\rho-\mu$ curve at density $\rho=1/2$ indicates the presence of an incompressible insulating phase where the superfluid density goes down to zero. 

As we tune the value of $\tau$ from $1$ to $0$, the nature of the insulator at half-filling goes through a phase transition, from a CDW phase to a dimer insulator (DI) phase. The structure factor $S(\pi,\pi)$, shown in Fig.~\ref{fig:phase_diagram}\,c, is finite only in the upper range of $\tau$ where the model approaches the isotropic $t-V$ model limit, indicative of a CDW. Indeed, in the case of a half-filled $t-V$ model, all particles simultaneously occupy one of the two sublattices in order to avoid repulsion, which results in an insulating CDW phase, with particles frozen in a single sublattice. However, when the anisotropy increases, the system crosses into a phase where the structure factor is zero while the dimer structure factor $S_D(\pi,\pi)$ peaks instead, thus ruling out the CDW nature of the insulator. 

Interestingly, by means of a simple analytical calculation we can actually point out the exact transition line for an infinite lattice. For $\tau=0$, the particles forming the dimers do not feel any repulsion and they can freely hop back and forth along the $\langle i,j\rangle_1$ bonds. Each particle gains an amount of energy $t$ while forming a dimer. Now, as soon as we tune $\tau$ to some nonzero value, the particles in the system start feeling repulsion from each other. As dimers are formed at each of the $\langle i,j\rangle_1$ bonds at any instant of time the constituent particle of a particular dimer can be repelled by its two neighboring sites along the $\langle i,j\rangle_2$ or $\langle i,j\rangle_3$ bonds. Since the average density of the sites are $0.5$, each of the HCBs will feel $2V^\prime\times 0.5=V^\prime$ amount of repulsion in presence of a nonzero value of $\tau$. Therefore, the hopping process in the formation of dimers will be preferred over the CDW pattern as long as we have $t>V^\prime$. In other words, the phase boundary between the topological dimer insulator and CDW will be at $t=V^\prime=\tau V$, setting the critical value of $\tau$ at $\tau_c=t/V$. The vertical dashed line in Fig.~\ref{fig:phase_diagram} indicates this transition line, in the thermodynamic limit, as obtained from the considerations above. In addition, we require $V>6t$ for the DI to be favorable over a superfluid phase in the extreme anisotropic limit, as each particle forming a dimer avoids $V/2$ repulsion while sacrificing $3t$ of kinetic energy.

Now, when $V^\prime$ is set to be zero, due to absence of repulsion along the $\langle i,j\rangle_2$ and $\langle i,j\rangle_3$ bonds it is energetically favorable for the particles to hop back and forth along the $\langle i,j\rangle_1$ bonds in order to further lower the energy of the system. Consequently, dimers are formed along all the $\langle i,j\rangle_1$ NN bonds in the system giving rise to a peak in $S_D(\pi,\pi)$ with maximum possible amplitude $1$. In this situation there is no net flow of HCBs in $x$ or $y$-direction of the bulk of the lattice which makes the phase insulating in nature. This however changes at the edge: it turns out that the DI is in fact an interacting topological insulator, as we now turn to describe. In the rest of the paper we identify its topological properties, starting from employing a bulk measure for topological order.


\begin{figure*}[t]
	\centering
	\includegraphics[width=0.30\textwidth]{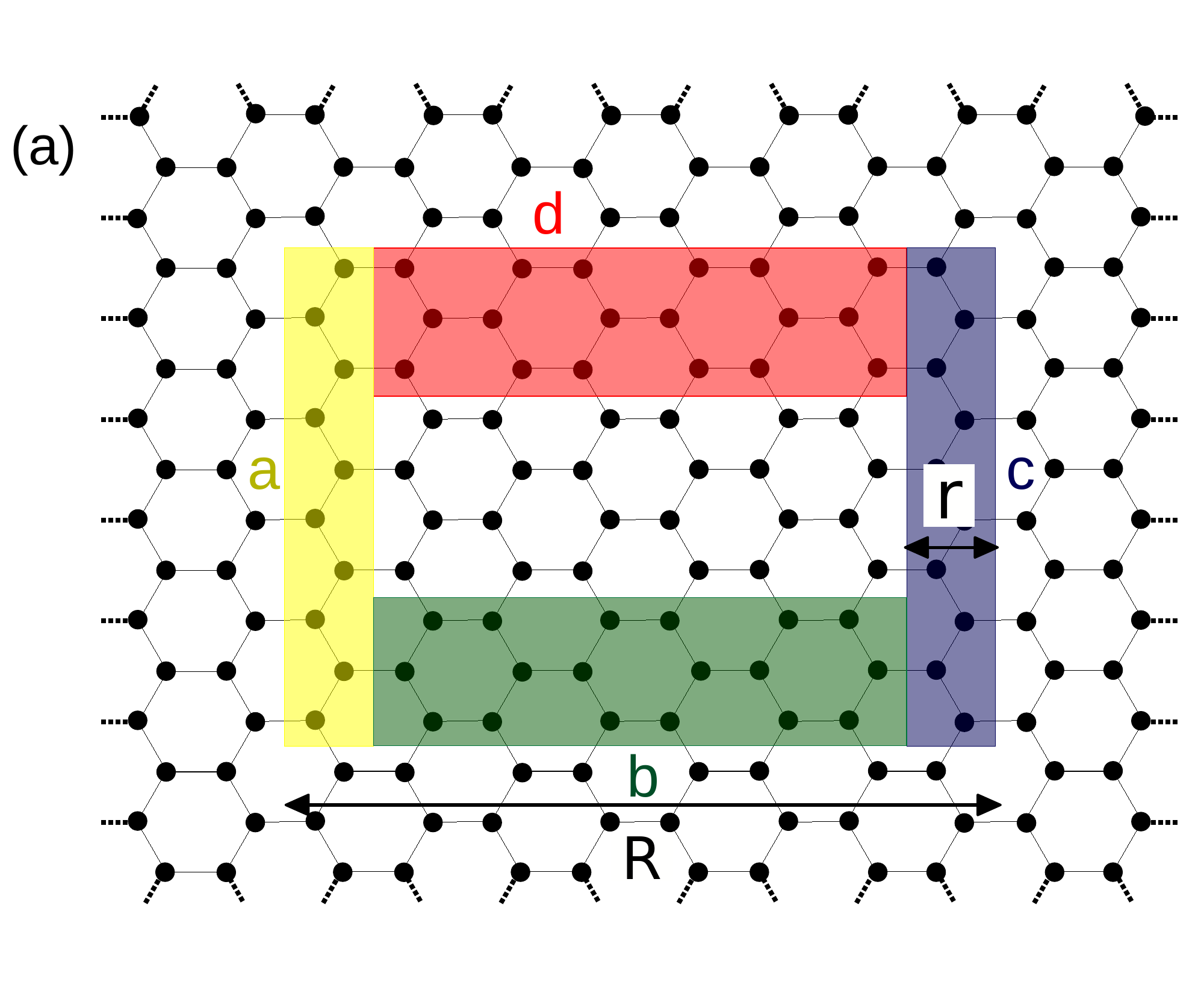}
	\includegraphics[width=0.33\textwidth]{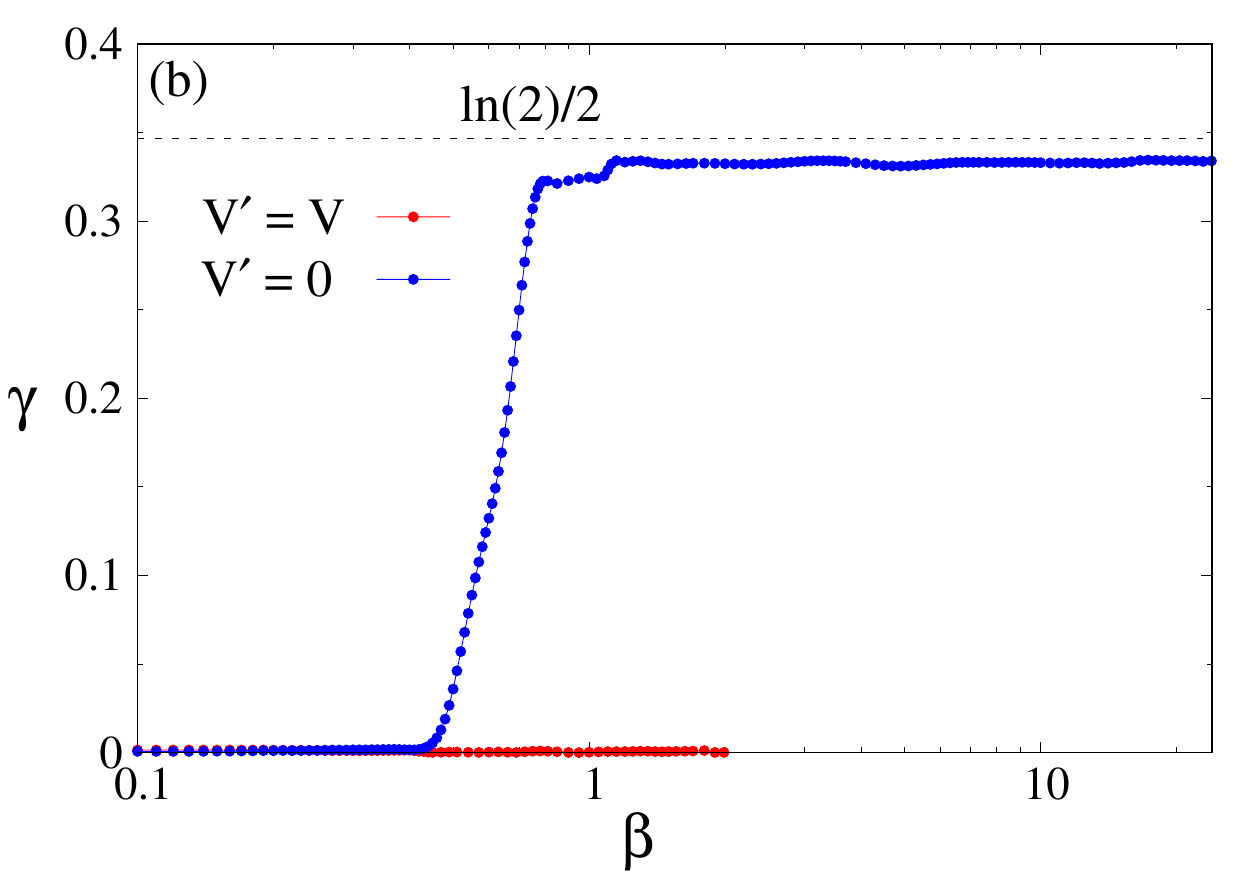}
	\includegraphics[width=0.33\textwidth]{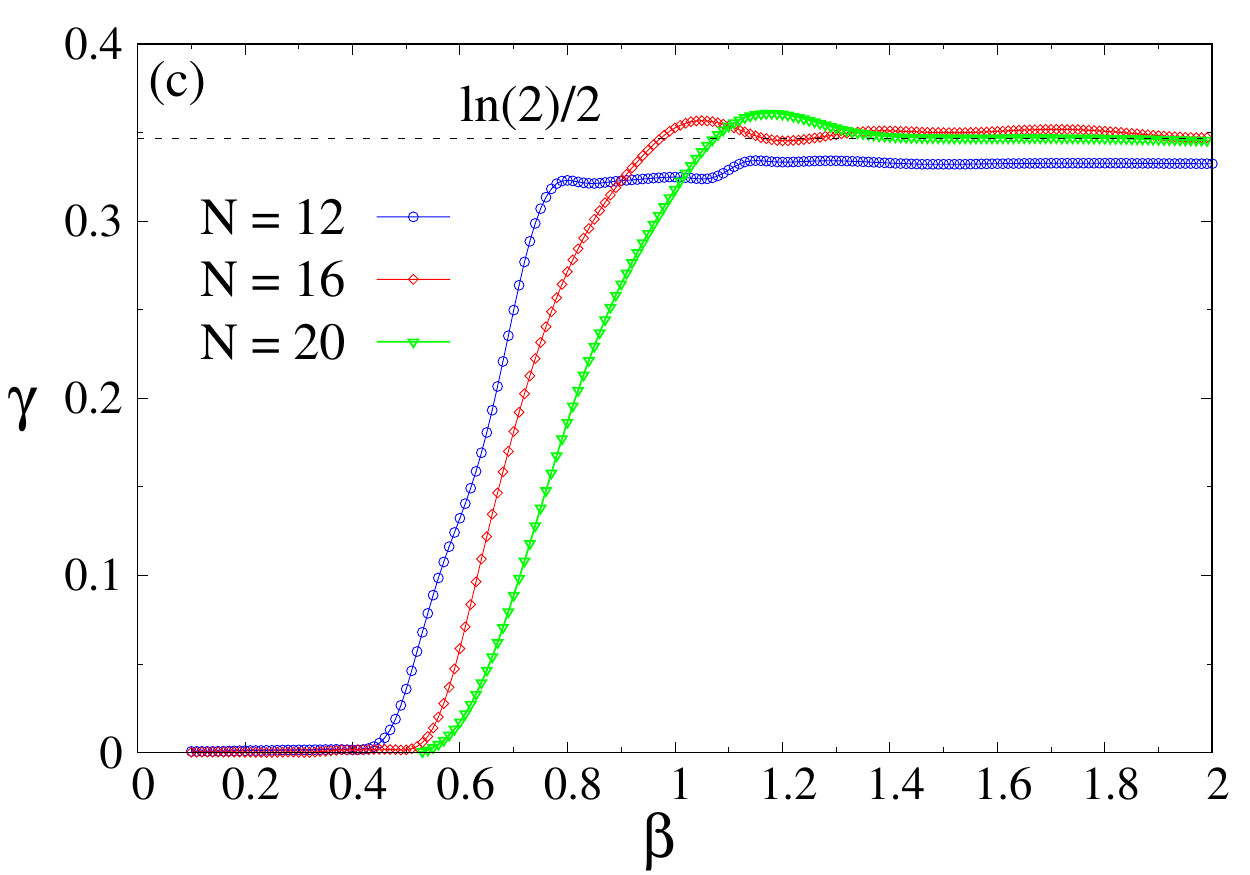}
	\caption{Topological entanglement entropy for the model described by Eq.~\eqref{eq:hamiltonian} with periodic boundary conditions. (a) Levin and Wen’s construction of the four subsystems, $A_1=a\cup b \cup c \cup d$, $A_2=a\cup b\cup c$, $A_3=a\cup d\cup c$ and $A_4=a\cup c$, used in the calculation of the topological entanglement entropy (schematically). (b) Topological entanglement entropy versus inverse temperature $\beta$ measured on half-filled $12\times 12$ honeycomb lattice for $V=8$ with both $V^\prime=V$ and $V^\prime=0$. (c) The same for three different system sizes, $12\times 12$, $16\times 16$ and $20\times 20$ at half-filling with $V=8$ and $V^\prime=0$.}
	\label{fig:topological_EE}
\end{figure*}

\emph{Topological entanglement entropy.---} Based on the idea that topologically ordered states contain non-local entanglement in the ground-state wavefunction, the topological entanglement entropy (TEE) was identified \cite{hamma2005ground,hamma2005bipartite,kitaev2006topological,levin2006detecting} as a quantity that can detect these correlations which are not revealed by the traditional long range orders. This quantity can be thought of as an order parameter which is nonzero in a topologically ordered state and zero otherwise. Now, the $n^{\text{th}}$ Renyi entanglement entropy between a subsystem $A$ and its complement $B$ (such that $A\cup B$ represents the whole system) is given by the formula
 $S_n(A)=\frac{1}{1-n}\ln\left[\mathrm {Tr}\left(\rho_A^n\right)\right],$
with $\rho_A$ being the reduced density matrix of subsystem $A$. The non-local correlations in a topologically non-trivial phase give rise to an area-law correction of the Renyi entanglement entropy, which in a 2D system reads as
$ S_n(A)=a L-j \gamma$.
In this equation, $a$ represents a non-universal constant, $L$ is the boundary length between subsystem $A$ and its complement, and $j$ gives the number of connected components in the subsystem $A$. Levin and Wen \cite{levin2006detecting} showed that by measuring the entanglement entropy for the four subsystems as depicted in Fig.~\ref{fig:topological_EE}\,a, one can extract the topological component $\gamma$ as,
\begin{align}
 2\gamma=\lim_{r,R\to\infty} \left[-S_n(A_1)+S_n(A_2)+S_n(A_3)-S_n(A_4)\right].
\end{align}
Since the ground-state wavefunction is inaccessible through QMC techniques, previously the calculation of the TEE was limited to either analytically solvable models or to small system sizes pertained to exact diagonalization method. More recently, Melko \emph{et~al.} \cite{melko2010finite} developed a QMC method based on the replica trick to calculate the Renyi entanglement entropy $S_n(A)$ for $n\geq 2$ at non-zero temperatures, which paved the way for calculating TEE in larger systems \cite{isakov2011topological}.

Fig.~\ref{fig:topological_EE}\,b depicts the variation of TEE as a function of the dimensionless inverse temperature $\beta=t/T$ measured on a half-filled $12\times 12$ honeycomb lattice for both isotropic ($V^\prime=V$) and anisotropic ($V^\prime=0$) interactions with $V=8$. We can see that in case of the isotropic $t-V$ model, the TEE remains zero throughout revealing the topologically trivial nature of the CDW. On the other hand, in the extreme anisotropic limit ($V^\prime=0$), as $T\to 0$ the TEE approaches a plateau with a value very close to $\ln(2)/2$. Fig.~\ref{fig:topological_EE}\,c compares the anisotropic case for three different system sizes, $12\times 12$, $16\times 16$ and $20\times 20$. For larger system size, the plateau appears to be more accurately quantized at $\ln (2)/2$. If $D$ represents the total quantum dimension of a state, then $\gamma = \ln D$. The quantum dimension takes a value $D > 1$ in a topologically ordered state which gives rise to a nonzero value of $\gamma$, whereas topologically trivial states (or certain non-interacting topological phases) have $D=1$. In Fig.~\ref{fig:topological_EE}\,c, since the plateau in $\gamma$ is quantized at $\ln(2)/2$, the quantum dimension of the anisotropic $t-V$ model turns out to be $\sqrt{2}$. This proves that the dimer insulator at half-filling is indeed a topological insulator, and is consistent with fractional QHE for bosons at filling factor $1/2$.


\begin{figure*}[t]
	\centering
	\includegraphics[width=0.33\textwidth]{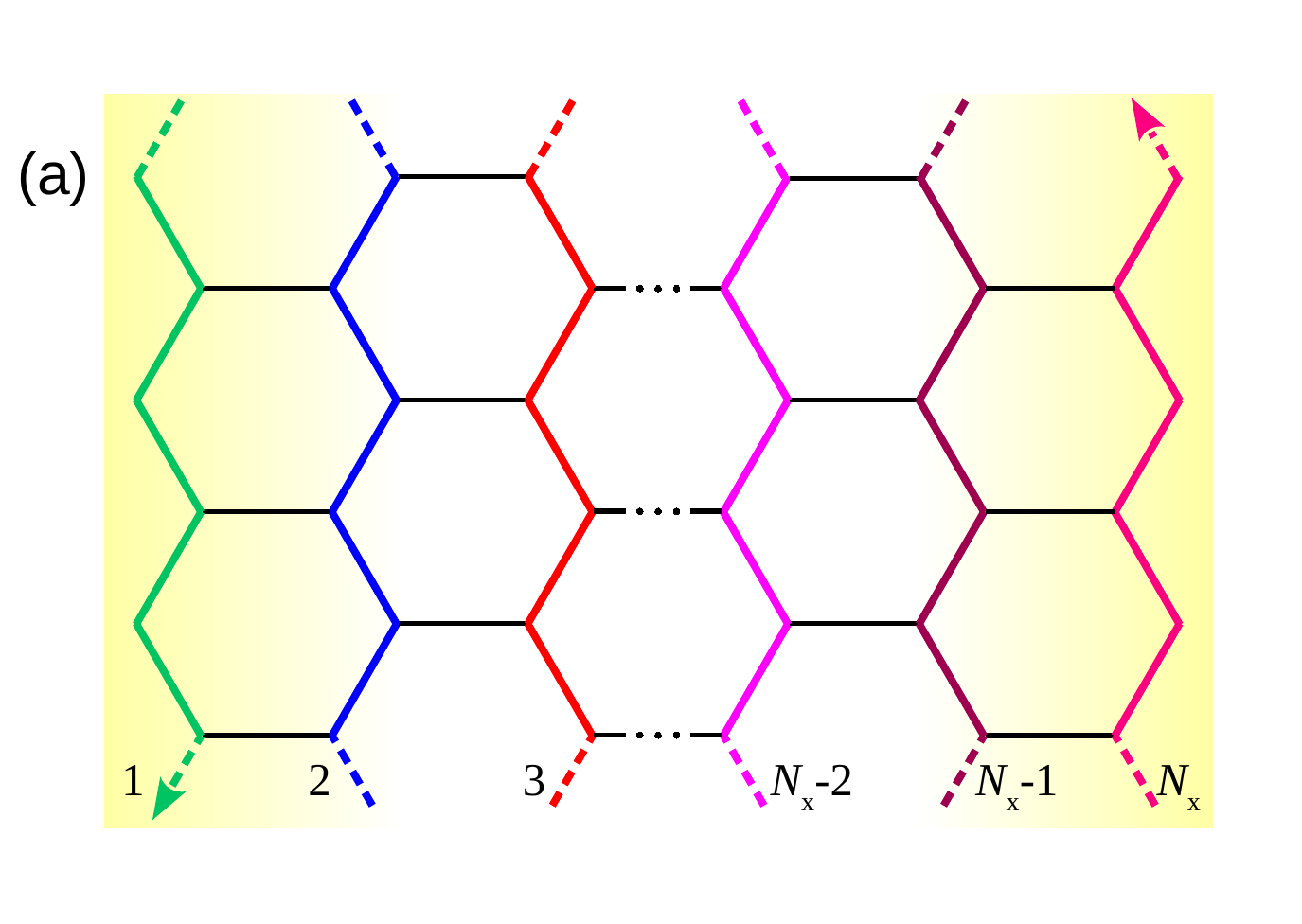}
	\includegraphics[width=0.31\textwidth]{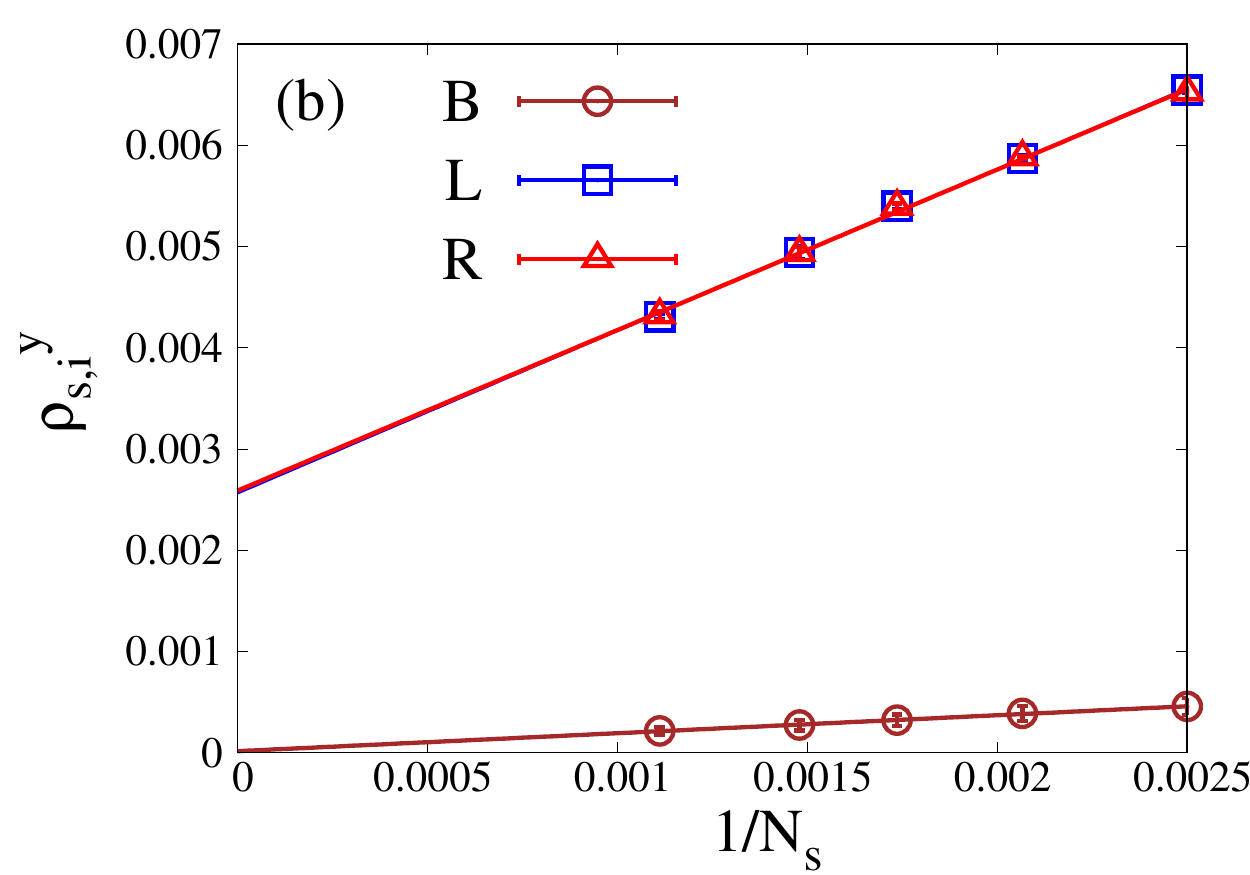}
	\includegraphics[width=0.31\textwidth]{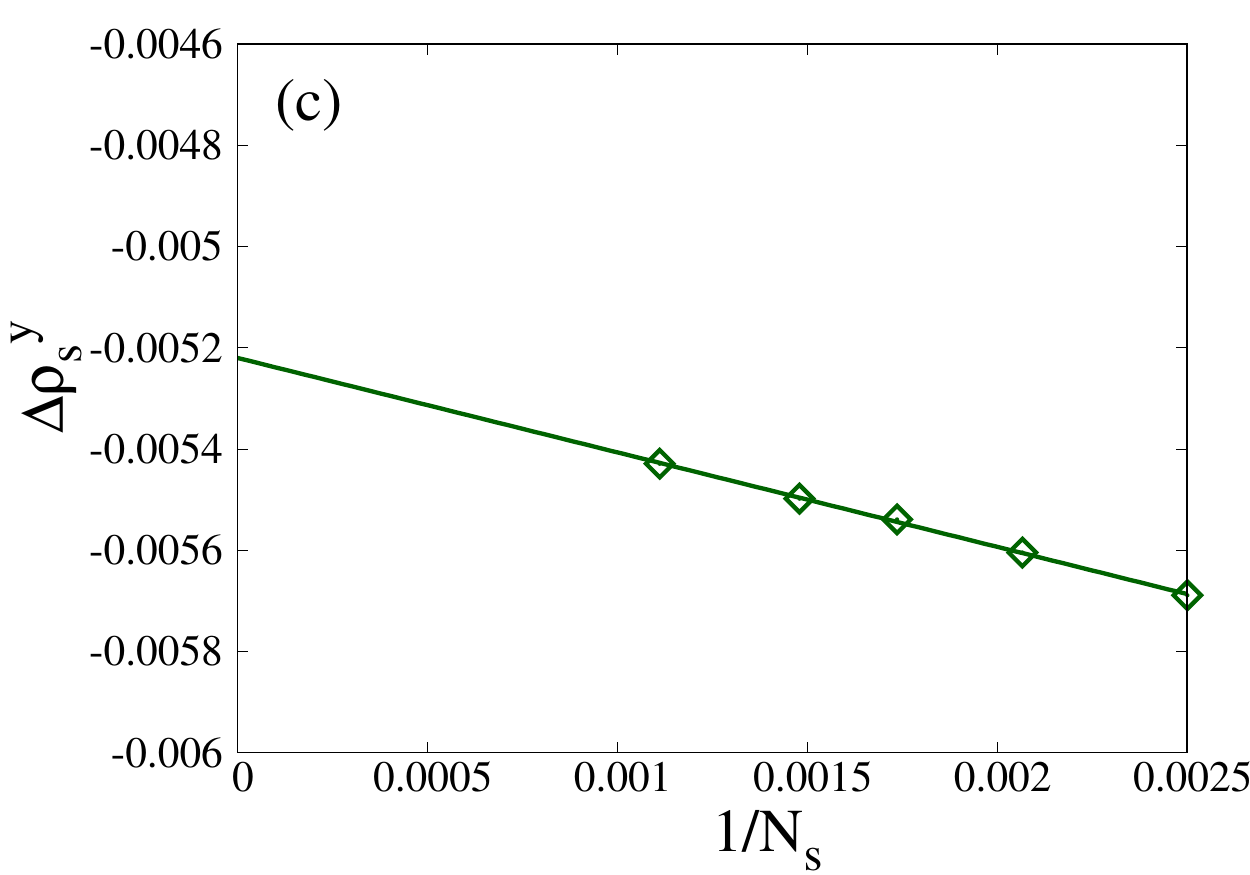}
	\caption{Detection of edge states and chirality in the DI phase. (a) An illustration of the stripes along which the superfluid density is measured, with regions of finite $y$-directed superfluid density highlighted in yellow. 
(b) System size dependence of the superfluid density for a stripe in the bulk (B), at the left edge (L) and at the right edge (R) of the system; and, (c) the same for $\Delta\rho_s^y$ in Eq.~\eqref{eq:delta_rho}.}
	\label{fig:chirality}
\end{figure*}

\emph{Edge states and chirality.---} Next, we employ a method for detecting the presence of edge states using QMC. We perform a switch of the boundary condition from periodic to open along a cut parallel to the $y$-axis, and study the changes in the density variations. For example, we observe that cutting a $20\times 20$ periodic honeycomb lattice along the $y$-axis, the plateau at $\rho=1/2$ splits into two plateaus corresponding to densities $\rho_1=0.475$ and $\rho_2=0.525$ (inset of Fig. \ref{fig:phase_diagram}\,b). Further study of this splitting on different system sizes reveals that for a $N_x\times N_y$ honeycomb lattice, the splitting densities can be expressed as $\rho_1=\rho-\frac{1}{N_x}$ and $\rho_2=\rho+\frac{1}{N_x}$. The origin of this splitting can be understood by realizing that, in case of a topological phase, the bulk-boundary correspondence implies that under open boundary condition we must have degenerate in-gap boundary states. Due to the appearance of these in-gap edge states, in the $\rho-\mu$ variation under open boundary condition one should obtain two plateaus: one corresponding to the situation when none of the edge sites are occupied and another one when all of the edge sites are completely occupied beyond some critical chemical potential. Study of the local density profile of the lattice at densities $\rho_1$ and $\rho_2$ demonstrates that, the splitting of the half-filling plateau in Fig. \ref{fig:phase_diagram}\,b in fact signifies such a situation. Indeed, in Fig. \ref{fig:phase_diagram}\,b, $(\rho_2-\rho_1)N_s=20$ which is exactly equal to the number of edge sites in a $20\times 20$ honeycomb lattice with zigzag edges. Interestingly, the edge states do not appear for a cut along a line parallel to the $x$-axis, \emph{i.e.}, the direction parallel to the underlying dimers, which would generate armchair edges but would not cut dimers.

To probe the chiral nature of these anisotropic edge states, we divide the open honeycomb lattice into zigzag stripes along the $y$-direction, see Fig.~\ref{fig:chirality}\,a. 
Fig.~\ref{fig:chirality}\,b demonstrates how the superfluid density along a stripe in the bulk of the system vanishes in the thermodynamic limit, while for both of the edge stripes it converges to the same non-zero value. This implies that for a lattice in the thermodynamic limit the bulk behaves as an insulator, whereas the edges remain conducting with the same value of superfluid density. 

Now, since superfluidity is related to the square of the winding number $(W^y)$, it does not carry the signature of the direction of the current. Instead, the direction can be captured from the sign of the winding number itself. However, for a finite lattice the current at the edges can always switch its direction during different measurements (since both are equally probable), which leads to a zero winding number. Therefore, to determine the chirality of the edge currents we form the quantity
\begin{eqnarray}
	\Delta\rho^y_s=\rho_s^y-\sum_{i=1}^{N_x}\rho^y_{s,i}, \label{eq:delta_rho}
\end{eqnarray}
where $\rho^y_s$ is the total superfluid density along the $y$-direction and $\rho_{s,i}^y$ is the superfluid density of the zigzag stripes of Fig.~\ref{fig:chirality}\,a. We claim that a measurement of a negative value for this quantity indicates chirality in our system, under certain consistency conditions as we now explain.
 
Indeed, $\rho^y_s$ and $\rho_{s,i}^y$ are related via the winding numbers along the $y$-direction of the different stripes, $W^y_i$, via,
\begin{align}
	\rho_s^y=\sum_i\rho_{s,i}^y+\frac{2}{\beta}\sum_{i\neq j}\langle W^y_i W^y_j\rangle.
\end{align}
For chiral edge states, with equal and opposite current persisting along the edges, one must have $\langle W^y_{\mathrm{L}}\rangle=-\langle W^y_{\mathrm{R}}\rangle$ (where L and R stand for left edge and right edge, respectively) with $\langle W^y_i\rangle=0$ for all other bulk stripes, therefore
$\Delta\rho^y_s\simeq -2\rho^y_{s,\mathrm E},\label{delta_rho}$ where $\rho^y_{s,\mathrm E}$ is the average superfluidity of the edges. We note that the above analysis holds only for an infinite lattice, since numerically there will always be a small bulk-current for a finite lattice. Fig.~\ref{fig:chirality}\,c depicts the dependence of $\Delta\rho_s^y$ on different system sizes ranging from $20\times 20$ to $30\times 30$, where in the thermodynamic limit it converges to a negative value very close to $-2\rho_{s,\mathrm E}^y$ as measured in Fig.~\ref{fig:chirality}\,b. This demonstrates that the edge states in our system are indeed chiral in nature.


\emph{Conclusions.--} We have characterized the phase diagram of an anisotropic version of the $t-V$ model, where the anisotropy is present in the NN repulsive interactions. We demonstrated that the CDW phase at half filling of the $t-V$ model is transformed to a topological DI at large anisotropies, that admits a finite TEE and, for open boundary conditions in the direction set by the dimers, admits chiral edge states. This phase is one of the simplest realization that we are aware of for the bosonic fractional quantum Hall effect, and arises, remarkably, in the absence of magnetic flux or lattice frustration. Instead, it employs anisotropy in the interaction similarly to Kitaev's honeycomb model \cite{kitaev2006anyons} but realizing a distinct state with a different value of the TEE. The generated state appears to be an anisotropic fractional quantum Hall state, similar to the one generated by the wire construction of the QHE \cite{kane2002fractional}, but with more natural types of interactions. The model and its phases could in principle be realized using cold atoms.


\begin{acknowledgments}
	We thank M.~Hastings and R.~Melko for discussions. This research was funded by the Israel Innovation Authority under the Kamin program as part of the QuantERA project InterPol, and by the Israel Science Foundation under grant 1626/16. AG thanks the Kreitman School of Advanced Graduate Studies for support. AG also thanks M. Sarkar and A. Basu for discussions.
\end{acknowledgments}

\bibliography{bibliography}

\end{document}